\documentclass[aps,amsfonts,showpacs,nofootinbib,twocolumn]{revtex4}

\usepackage{graphicx}
\usepackage{dcolumn}
\usepackage{bm}

\newcommand{\be}{\begin{equation}}
\newcommand{\ee}{\end{equation}}
\newcommand{\ba}{\begin{eqnarray}}
\newcommand{\ea}{\end{eqnarray}}
\def\beu{\begin{displaymath}}
\def\eeu{\end{displaymath}}

\begin{document}

\title{The optimal approach of detecting stochastic gravitational wave from string cosmology using multiple detectors}

\author{Xi-Long Fan and Zong-Hong Zhu\footnote{zhuzh@bnu.edu.cn}}

\address {Department of Astronomy, Beijing Normal University,
Beijing 100875, People's Republic of China}

\date{\today}
\begin{abstract}
String cosmology models predict a relic background of gravitational
wave produced during the dilaton-driven inflation. It's spectrum is
most likely to be detected by ground gravitational wave laser
interferometers (IFOs), like LIGO, Virgo, GEO, as the energy density
grows rapidly with frequency. We show the certain ranges of the
parameters that underlying string cosmology model using two
approaches, associated with $5\%$ false alarm and $95\%$ detection
rate. The result presents that the approach of combining multiple
pairs of IFOs is better than the approach of directly combining the
outputs of multiple IFOs for LIGOH, LIGOL, Virgo and GEO.
\end{abstract}


\maketitle

\section{Introduction}
Stochastic gravitational wave background, which has two origins, is
one target of gravitational wave interferometers (IFOs). It might
result from an extremely large number of weak astrophysical
gravitational wave sources, like compact stars in binary system (see
e.g.\cite{binary background} for more details). It also might result
from some processes of very early universe, like phase transitions
or amplification of vacuum fluctuations in inflationary (see
e.g.\cite{maggiore:2000,Allen:lec,buonanno} for reviews). In the
latter case of origin, the gravitational waves carry the earlier
information of the universe than that shown by electromagnetic
waves. One of the most interesting processes in the early universe
is from the string cosmology\cite{string_1,string_2}, which predicts
a quite different gravitational wave background spectrum from that
predicted by other  cosmological models for early universe. That the
energy density grows rapidly with frequency\cite{spe_grow} means
that the ground IFOs may be the best detectors. Several large scale
ground IFOs  are in operation: Laser Interferometric Gravitational
Wave Observatory (LIGO)\cite{LIGO} in Livingston (LIGOL) and in
Hanford(LIGOH), Virgo\cite{VIRGO} near Pisa and GEO\cite{GEO} in
Hanover.

\begin{figure}
\includegraphics[width=3.3in]{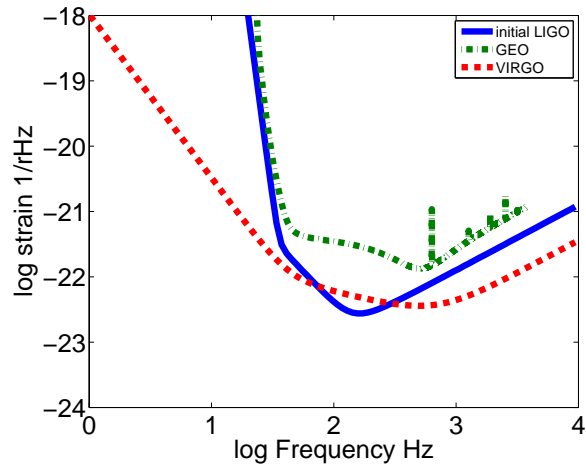}
\caption{\label{s_f} The designed noise power spectrum of initial
LIGO, Virgo and GEO. Data are taken
from\cite{ligo_noise,virgo_noise,geo_noise}}
\end{figure}

\begin{figure}
\includegraphics[width=3.3in]{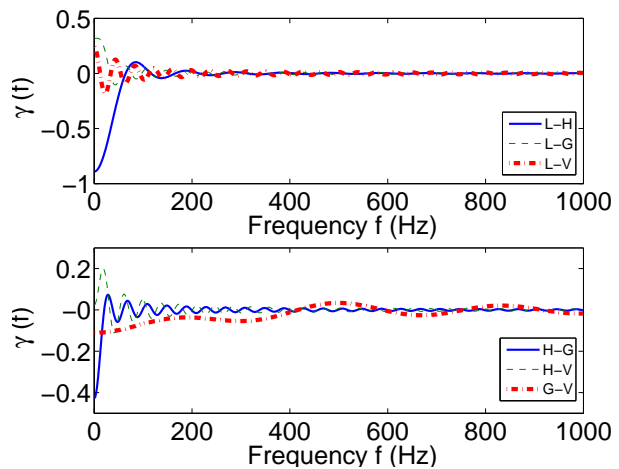}
\caption{ The overlap reduction function. Data are taken
from\cite{ligo_noise,virgo_noise,geo_noise,grasp}} \label{g_f}
\end{figure}

Two approaches of combining $2N$ detectors to improve the detection
ability to the stochastic gravitational wave back ground are
proposed in\cite{Allen:1999}: (i)correlating the outputs of a pair
of detectors, then combing the multiple pairs, and (ii) directly
combing the outputs of $2N$ detectors. As shown
in\cite{n_detectors}, for $2N$ detectors with equal noise level, the
data observation time and the overlap functions, the optimal
approach is to combine multiple pairs of two detectors comparing to
directly combing $2M (M\leq N)$ detectors. But the real detectors
should not be of identical noise levels and overlap functions. We
plot the proposed noise curves of detectors in Fig. \ref{s_f} and
the overlap reduction functions in Fig. \ref{g_f}. A number of
authors\cite{Allen:1997,RB98,MB,MS} have used the approach of
combining LIGOH  and LIGOL to detect the string cosmology
gravitational wave background. A recent work\cite{V_L} has shown
that the approach of combining multiple pairs of IFOs using Virgo
and  LIGO and GEO can improve the detection ability to the
stochastic gravitational wave background illustrated by  a simulated
isotropic gravitational wave background generated with an
astrophysically-motivated spectral shape.

In this paper we compare the two approaches by constraining the
parameter space of gravitational wave background predicted from
string cosmology using LIGOH, LIGOL, Virgo, and GEO. Our result
shows that the approach of combining multiple pairs of IFOs is
better than the approach of directly combining the outputs of
multiple IFOs for those real IFOs at their designed noise levels.
Our paper is organized as follows. In Sec. \ref{two approaches} we
review the two approaches of detecting a stochastic background using
multiple detectors. In Sec. \ref{implement} after a brief review of
the gravitational wave background produced by string cosmology, we
implement the two approaches using four IFOs: LIGOH, LIGOL, Virgo
and GEO at their designed noise levels. Our conclusion will be
provided in Sec. \ref{conclusion} shows.

\section{Two approaches of detecting a stochastic background using multiple
detectors}\label{two approaches}

 It has been shown\cite{Allen:1999} that after correlating signals of two detectors for time $T$ (we take
$T=10^7 \> {\rm sec} = 3 \> {\rm  months}$) the squared ratio of
``Signal"  ($S$) to ``Noise" ($N$) is given by an integral over
frequency $f$: \be\label{snr}
 \left( {S \over N} \right)^2 = {9 H_0^4 \over 50 \pi^4} T
\int_0^\infty df \> {\gamma^2 (f) \Omega_{\rm gw}^2(f) \over f^6
P_1(f) P_2(f)}\ ,
 \ee
 where $P_i(f)$ is the one-side noise power spectral density
 which describes the instrument noise $n_i (f)$ in frequency domain:
\be \langle\tilde n_i^*(f)\tilde n_i(f')\rangle= {1\over
2}\delta(f-f')\ P_i(|f|)\ . \ee

Eq.~\ref{snr} is under the assumption that the  noise of the
detectors are (i) stationary, (ii) Gaussian, (iii) statistically
independent of one another and of the stochastic gravitational wave
background, and (iv) much larger in magnitude than the stochastic
gravitational wave background.

 $\gamma(f)$ is the so-called
{\it overlap reduction function} first calculated by
Flanagan\cite{flan}, which shows the co-response of two detectors.
This is a dimensionless function of frequency $f$, which is
determined by the relative positions and orientations of two
detectors. Explicitly,

\begin{equation}\label{gamma_f}
\gamma(f):={5\over 8\pi}\sum_A\int_{S^2}d\hat\Omega\ e^{i2\pi
f\hat\Omega\cdot\Delta \vec x/c}\
F_1^A(\hat\Omega)F_2^A(\hat\Omega)\},
\end{equation}
where $\hat \Omega$ is a unit vector specifying a direction on the
two-sphere, $\Delta\vec x:=\vec x_1-\vec x_2$ is the separation
vector between the central stations of the two detector sites, and
\begin{equation}\label{e:F_i^A}
F_i^A(\hat\Omega):= e_{ab}^A(\hat\Omega)\ {1\over 2}\left(\hat
X_i^a\hat X_i^b- \hat Y_i^a\hat Y_i^b\right)\
\end{equation}
is the $i_{th}$ detector's response to a zero frequency, unit
amplitude, $A = + ,\times$ polarized gravitational wave, where $\hat
X_i^a $ and  $\hat Y_i^a$ are unit vectors pointing in the direction
of the detector arms. The overlap reduction function $\gamma(f)$ in
 Eq.~\ref{gamma_f} is normalized for coincident and coaligned detectors: $\gamma(0)=1$. we refer the reader to \cite{Allen:1999,maggiore:2000} for
more details about the overlap reduction function $\gamma(f)$. Two
approaches were shown in \cite{Allen:1999} for multiple IFOs, the
optimal approach of combing multiple detector pairs:
\be\label{snr_pair_opt}
 \left( {S \over N} \right)^2_{optI} = \sum_{pair}\left( {S \over N} \right)^2_{pair}
 ,
 \ee
 and the optimal approach of directly combing $2N$ detectors:
 \begin{equation}\label{snr_2n_opt}
\begin{array}{cc}
\left( {S \over N} \right)^2_{op\,tII}\approx & {}^{(12)} \left( {S \over N} \right)^2\ {}^{(34)} \left( {S \over N} \right)^2 \cdots{}^{(2N-1,2N)} \left( {S \over N} \right)^2 \\
   &  +\hbox{{\rm\ all possible
permutations}}\ .
\end{array}
\end{equation}

 In order to detect a stochastic background with $5\%$ false alarm and $95\%$ detection
rate, the total optimal signal to noise ratio  $SNR_{opt}$ threshold
should be $3.29$.

\begin{figure}
\includegraphics[width=3.3in]{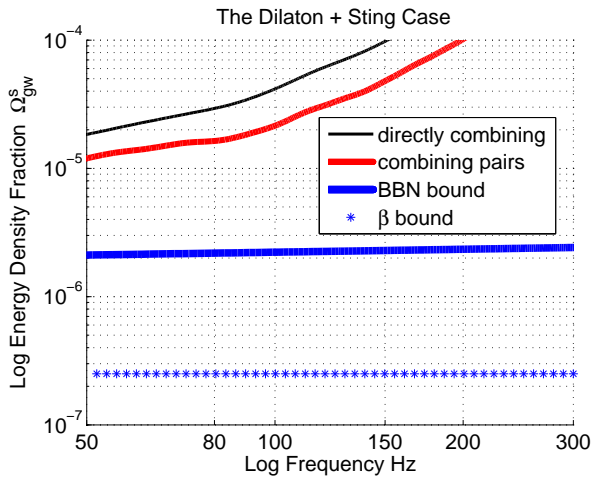}
\caption{ The region of parameter space for which the gravitational
wave stochastic produced in the ``dilaton + string" case is
detectable by the approach of combining multiple pairs of IFOs
(labeled ``combine pairs") and the approach of directly combining
four IFOs (labeled ``directly combine"). With $5\%$ false alarm and
$95\%$ detection rate, the region above of the curves shows the
detectable parameter space of the stochastic gravitational wave
background produced in the ``dilaton + string" case by networks of
detectors. It is clear that the ``combine pairs" approach has higher
detection ability for the gravitational wave background. But the
``combine pairs" approach is still far away from observing the
stochastic gravitational wave background spectrum. }\label{string}
\end{figure}

\begin{figure}
\includegraphics[width=3.3in]{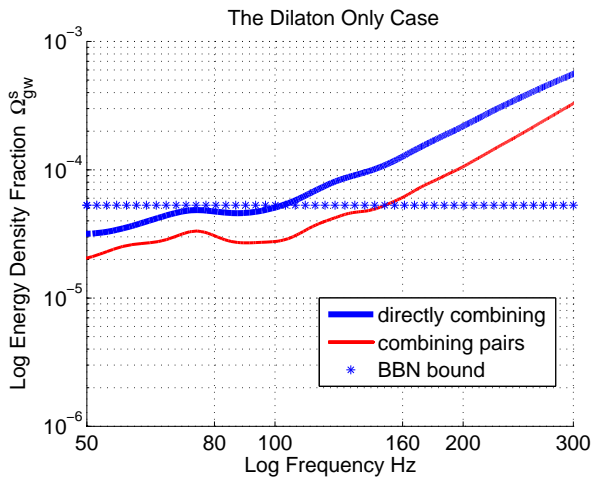}
\caption{The region in parameter space for which the gravitational
wave stochastic background is observable by two approaches with
$5\%$ false alarm and $95\%$ detection rate. The region above of the
curves shows the detectable parameter space of the stochastic
gravitational wave background produced in the ``dilaton only" case
by networks of detectors. It is clear that the ``combine pairs"
approach has more chance to detect the gravitational wave
background.}\label{only}
\end{figure}

\section{Detecting a string cosmology background using two approaches by four IFOs}\label{implement}

String cosmology, also denoted in the literature as the
``pre-Big-Bang (PBB) models" depicts a different view of PBB era
from the ``slow-roll" (standard) inflation. The dilaton-driven
inflation phase is well understood, followed by a sting phase which
is currently not known. The ``minimal" PBB model\cite{GMV} describes
the sting phase following the dilaton-driven inflation phase is a
constant curvature phase. At the end of the string phase, the
dilaton reaches the present vacuum expectation value and stops. The
spectrum of gravitational wave background predicted during the
periods of dilaton-driven inflation phase and string phase in string
cosmology is discussed in\cite{spectrum}. The simplest model, in
which the approximate form of the spectrum is\cite{RB96} \be
\label{approx} \Omega_{\rm gw}(f)= \cases{ {\Omega^{\rm s}_{\rm gw}
(f/{f_{\rm s}})^3}  &  { $f<{f_{\rm s}}$} \cr
 & \cr
{\Omega^{\rm s}_{\rm gw} (f/{f_{\rm s}})^{\beta} } &   ${f_{\rm
s}}<f<f_1$ \cr
 & \cr
{0} &   $f_1<f$}\ , \ee where \beu \beta=\frac{\log\left[\Omega_{\rm
gw}^{\rm max}/ \Omega^{\rm s}_{\rm gw}\right]}{\log\left[f_1/{f_{\rm
s}}\right]} \eeu is the logarithmic slope of the spectrum produced
in the string phase. The spectrum depends on four parameters: the
frequency ${f_{\rm s}}$ and the fractional energy density
$\Omega^{\rm s}_{\rm gw} $ produced at the end of the dilaton-driven
inflation phase, the maximal frequency $f_1$ above which
gravitational radiation is not produced and the maximum fractional
energy density $ \Omega^{\rm max}_{gw} $ which occurs at frequency
$f_1$. In this paper we follow\cite{Allen:1997} setting \be f_1= 1.3
\times 10^{10} \> {\rm Hz} \left( { H_{\rm r} \over 5 \times 10^{17}
\> {\rm GeV}} \right)^{1/2} \ee and \be \Omega_{\rm gw}^{\rm max} =
1 \times 10^{-7} {\rm h}_{100}^{-2} \left( { H_{\rm r} \over 5
\times 10^{17} \> {\rm GeV}} \right)^2, \ee assuming no late entropy
production and making reasonable choices for the number of effective
degrees of freedom. $H_{\rm r}$ is the Hubble parameter at the
string phase. The ``reduced" Hubble parameter $h_{100}$ is in the
range $0.4 \le h_{100} \le 0.85$ by observations.

By virtue of Eq.~\ref{snr_pair_opt} and Eq.~\ref{snr_2n_opt}, using
given multiple IFOs, for any given set of parameters we may
numerically evaluate the optimal signal to noise ratio $SNR_{opt}$;
if this value is greater than 3.29 then with $5\%$ false alarm and
$95\%$ detection rate, the background can be detected by those IFOs.
We compare the two approaches  for constraining the string cosmology
gravitational wave background space parameter adopting LIGOH, LIGOL,
Virgo and GEO at their designed noise levels\cite{ligo_noise,
virgo_noise, geo_noise}. The location information of the different
GW observatories were obtained from\cite{grasp} and references
therein. The regions of detection ability to parameters space for
the approach of combining multiple pairs of IFOs (labeled ``combine
pairs") and the approach of directly combining four IFOs (labeled
``directly combine") for spectrum in Eq.~\ref{approx} (labeled
``dilaton + string" case) are shown in Fig.~\ref{string}. We have
assumed ${\rm h}_{100}=0.63$\cite{h_100} and $H_{\rm r}=5 \times
10^{17} \> {\rm GeV}$.

The parameter $ \beta$ is determined by the basic physical
parameters of string cosmology models which are not well known. Just
to show the different detection abilities of two approaches, a
phenomenological model is adopted. This model is the ``dilaton only
case" in\cite{Allen:1997}, assuming NO stochastic background is
produced during the string phase of expansion, i.e.

\be \label{only_spe} \Omega_{\rm gw}(f)= \cases{ {\Omega^{\rm
s}_{\rm gw} (f/{f_{\rm s}})^3}  &  { $f<{f_{\rm s}}$} \cr
 & \cr
{0} &   $f_s<f$}\ . \ee It is phenomenologically interesting as it
is a model whose spectrum peaks in the real IFOs band.
Fig.~\ref{only} shows the regions of detection abilities to
parameter space for ``combine pairs" approach and ``directly
combine" approach.

We also plot the Big Bang nucleosynthesis (BBN)
bounds\cite{Allen:1997,bbn} to see the detective chance for the
assuming spectrum in Eq.~\ref{approx} and Eq.~\ref{only_spe}. Note
that in the ``dilaton + string" case, the  slope $\beta$ appearing
in Eq.~\ref{approx} must satisfy the constraint $ \beta \geq 0$ (see
e.g.\cite{book}). As a consequence, $\Omega_{gw}^{max} =
\Omega_{gw}(f_1)$ is always larger than $\Omega_{gw}^{s} =
\Omega_{gw}(f_s)$. So, there is another bound for $\Omega_{gw}^{s}$.
In this paper, by assuming ${\rm h}_{100}=0.63$ and $H_{\rm r}=5
\times 10^{17} \> {\rm GeV}$, $\Omega_{gw}(f_1) \simeq 0.25 \times
10^{-6}$ sets a tighter bound. We also plot this tighter bound
(labeled $``\beta"$ bound) in Fig.~\ref{string}.

\section{Conclusion} \label{conclusion}
Two optimal approaches of  combing multiple detectors to detect a
stochastic gravitational wave back ground are proposed in
\cite{Allen:1999}. As shown in Fig.~\ref{s_f} and Fig.~\ref{g_f},
the real detectors should not be of identical noise levels and
overlap functions. It is necessary to compare two optimal approaches
of detecting the stochastic gravitational wave back ground for any
given real IFOs. In this short paper, we have compared two
approaches of combing four real ground IFOs (LIGOH, LIGOL, Virgo,
GEO) to show the detection ability of the stochastic gravitational
wave background predicted by string cosmology of the early universe.
As shown in Fig.~\ref{string} and Fig.~\ref{only}, it is clear that
the approach of combing multiple pairs of IFOs shows more detective
chance to string cosmology in both ``dilaton + string" case
 and ``dilaton only" case than the approach of directly combing outputs of those four IFOs. In
the `` dilaton only" case, as shown in Fig.~\ref{only}, both
approaches have the chance to observe the spectral peak between 50
Hz and 100 Hz. The approach of combing multiple pairs of IFOs could
detect that background at a little higher frequency up to 160 Hz. In
the `` dilaton + string" case, both approaches are also far away
from observing the stochastic gravitational wave background
spectrum. We hope the next generation IFOs will constrain the
parameter tighter.

\section*{Acknowledgments}
XL. Fan thanks Xing Wu for careful reading manuscript. Our thanks go
to the anonymouse referee for valuable comments and useful
suggestions, which improved this work very much. This work was
supported by the National Natural Science Foundation of China, under
Grant No. 10533010, 973 Program No. 2007CB815401 and Program for New
Century Excellent Talents in University (NCET) of China.


\end{document}